\documentclass[twocolumn,showpacs,preprintnumbers,amsmath,amssymb,aip]{revtex4}

\usepackage{graphicx}
\usepackage{dcolumn}
\usepackage{bm}
\usepackage{color}

\begin{document}


\title{Single- and few-electron dynamic quantum dots in a perpendicular magnetic field}

\author{S.~J.~Wright$^{1,3}$, A.~L.~Thorn$^{1,2}$, M.~D.~Blumenthal$^{1,2}$, S.~P.~Giblin$^{2}$, M.~Pepper$^{4}$, T.~J.~B.~M.~Janssen$^{2}$, M.~Kataoka$^{2}$, J.~D.~Fletcher$^{2}$, G.~A.~C.~Jones$^{1}$, C.~A.~Nicoll$^{1}$, Godfrey~Gumbs$^{5}$, D.~A.~Ritchie$^{1}$\\
}

\affiliation{
${}^1$\footnotesize Cavendish Laboratory, University of Cambridge, J. J. Thomson Avenue, Cambridge CB3 0HE, UK.\\
${}^2$\footnotesize National Physical Laboratory, Hampton Road, Teddington TW11 0LW, UK.\\
${}^3$\footnotesize Toshiba Research Europe Ltd, Cambridge Research Laboratory, 208 Science Park, Milton Road, Cambridge CB4 0WE, UK.\\
${}^4$\footnotesize University College London, Torrington Place, London WC1E 7JE, UK.\\
${}^5$\footnotesize Department of Physics and Astronomy, Hunter College of the City University of New York, 695 Park Avenue, New York, New York 10065 USA.
}

\date{\today}

\begin{abstract}

We present experimental studies of the current pumped through a dynamic quantum dot over a wide range of magnetic fields. At low fields we observe repeatable structure indicating increased confinement of the electrons in the dynamic dot. At higher fields ($B>3\,$T), we observe structure which changes markedly from device to device suggesting that in this regime the transport is sensitive to local disorder. The results are significant for the development of dynamic quantum dot pumps as quantum standards of electrical current.



\end{abstract}

\pacs{Valid PACS appear here}
\maketitle

\section{introduction}

A quantized charge transport device can generate electrical current given by $I=nef$, where $f$ is the repetition frequency of an applied potential, $e$ is the electron charge and $n$ is the number of charges transported in one cycle. This type of device is of great interest to electrical metrologists because it could form the basis of a new definition of the SI base unit ampere, linking current to frequency via a defined value of the electron charge~\cite{milton07}. Pumps based on chains of metal-oxide tunnel barriers have been researched extensively, and have demonstrated pumping accuracy at the 10$^{-8}$ level required by metrological applications~\cite{keller96}. Unfortunately the time constant of the tunnel junctions limits the current in these devices to the level of a few pA. Recently, a new type of pump based on metal-oxide-superconductor barriers has demonstrated parallel scaling of 10 devices~\cite{maisi09}, but this device must be operated at finite bias voltage, thereby requiring stringent control of leakage currents if metrological accuracy is to be reached.


The semiconductor-based dynamic quantum dot (DyQD) pump, in contrast, can be operated at zero bias, and relatively high frequency~\cite{blumenthal07}. The DyQD pump, like earlier Surface Acoustic Wave (SAW)-based pumps~\cite{shilton96,kataoka09}, transports electrons between a source and drain lead by modulation of the electrostatic potential in a reduced-dimensional semiconductor system. In the SAW pumps the potential modulation is produced by a SAW launched from a tuned transducer, whereas in the DyQD pump the modulation signal is applied directly to one of the potential-defining gates. The DyQD pump avoids heating effects present in the SAW pumps~\cite{schneble06}, and can be driven at a wide range of frequencies. Under the application of a perpendicular magnetic field, the performance of the DyQD pump was shown to be enhanced~\cite{wright08,kaestnerb}. Recent measurements at $B=5\,$T and $f=340\,$MHz did not resolve any error in the pump current within the 15 parts per million uncertainty in the current measurement system~\cite{giblin10}. Furthermore, parallel operation of two pumps has been demonstrated with no noticeable loss of accuracy~\cite{wright09}. The DyQD pump is therefore a strong candidate for the realization of a quantum standard of electrical current.

In this paper, we describe the effect of a perpendicular magnetic field on the current produced by a DyQD pump. For fields of $B\leq3\,$T the pumps exhibit phenomena that are reproducible from device to device. The risers between plateaus become sharper and the plateaus become flatter, indicating enhanced quantization. Transitions in the number of electrons transported per cycle shift in gate voltage, demonstrating the ability of the field to act as an extra control parameter to tune the pump system. At fields of $B>5\,$T an anomalous structure is observed in the quantized current that is reminiscent of earlier single-electron capacitance spectroscopy (SECS) measurements with static quantum dots (QDs)~\cite{ashoori1,zhitenev97}. The observation that the details of this structure are device-dependent suggests that they originate from local disorder which is unique to each device. The magnetic field appears to strengthen the effect of disorder on the measured pumped current.

\section{tunable-barrier electron pump}

\begin{figure}
\includegraphics[width=0.3\textwidth]{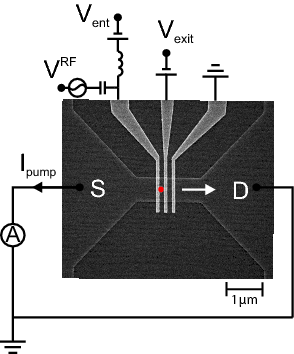}
\caption{\label{fig:device} SEM image of the device and schematic of electrical connections. The oscillating voltage signal $V^{\mathrm{RF}}$ is added to the static DC voltage $V_{\mathrm{ent}}$ and applied to the left (entrance) gate. $V_{\mathrm{exit}}$ is applied to the middle (exit) gate. The right gate is grounded and not used. A DyQD is periodically formed in the channel at the position of the red dot. Electrons are transported by the DyQD from source (S) to drain (D) reservoirs in the direction of the white arrow.}
\end{figure}

The DyQD pump devices are fabricated in a GaAs/AlGaAs high electron mobility transistor (HEMT) heterostructure where a two-dimensional electron gas (2DEG) exists 90$\,$nm below the surface. A scanning electron microscope (SEM) image of a similar device to the ones tested in this work is presented in Fig.~\ref{fig:device}. Ohmic contacts were made to the source (S) and drain (D) areas of 2DEG. Transverse confinement was provided by the horizontal narrow channel, created through shallow wet chemical etching. Metallic gates were deposited on the surface of the device, perpendicular to the channel. The left-most gate will be referred to hereafter as the \emph{entrance gate}, and the middle gate as the \emph{exit gate}. The right-most gate was grounded and not used. A sinusoidal radio frequency (RF) voltage signal $V^{\mathrm{RF}}$ was added to the static DC offset voltage $V_{\mathrm{ent}}$ using a bias tee, as shown, resulting in a total instantaneous entrance gate voltage $V_{\mathrm{ent}}^{\mathrm{TOT}}$. When tuned correctly, a DyQD is periodically formed at the position of the red dot in Fig.~\ref{fig:device} at the repetition frequency of $V^{\mathrm{RF}}$. A well-defined number of electrons can be captured by the DyQD from the source. As the pump cycle progresses and the potential is tilted, a controlled number of the captured electrons are ejected over the exit gate and into the drain, contributing to the measured current. The direction of electron transport is shown by the white arrow in the figure.

\begin{figure*}[t]
\includegraphics[width=0.8\textwidth]{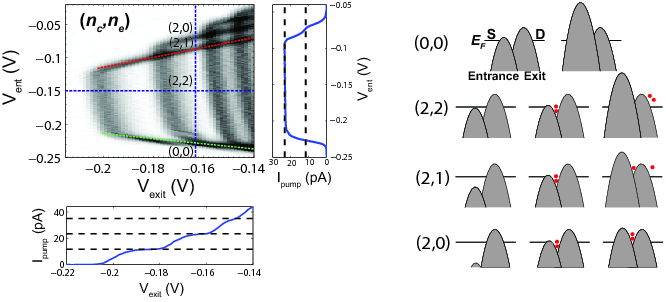}
\caption{\label{fig:pump_data} Left: the response of the pumped current to changes in $V_{\mathrm{ent}}$ and $V_{\mathrm{exit}}$. The main panel shows the numerical derivative in $V_{\mathrm{exit}}$ and $V_{\mathrm{ent}}$ of the pumped current, highlighting transitions in the number of pumped electrons. The blue dashed lines show directions of line scans in $V_{\mathrm{ent}}$ and $V_{\mathrm{exit}}$, seen to the right and bottom of the main panel respectively. Dashed black lines correspond to the expected plateau values. Right: schematic diagrams of the barriers defined by the gates in each of the four $\left( n_{c},n_{e}\right)$ regions, where $n_{c}$ is the number of electrons captured by the DyQD and $n_{e}$ is the number ejected into the drain.}
\end{figure*}

A plot of the numerical derivative of the pumped current in $V_{\mathrm{exit}}$ and $V_{\mathrm{ent}}$,
\begin{equation*}
\sqrt{\left(\frac{\mathrm{d}I_{\mathrm{pump}}}{\mathrm{d}V_{\mathrm{exit}}}\right)^{2}+\left(\frac{\mathrm{d}I_{\mathrm{pump}}}{\mathrm{d}V_{\mathrm{ent}}}\right)^{2}},
\end{equation*}
is presented in the main left panel of Fig.~\ref{fig:pump_data}. Here, $V^{\mathrm{RF}}$ was set to a frequency of $f=73\,$MHz with an amplitude at the source of $-9\,$dBm. All measurements in this work were performed in a dilution refrigerator with a base temperature of $\sim50\,$mK. Transitions in the number of electrons transported per cycle manifest in dark lines in the plot. We will refer to this type of plot as a \emph{pump map} hereafter. The blue dashed lines mark directions of line scans in $V_{\mathrm{ent}}$ and $V_{\mathrm{exit}}$, seen to the right and bottom of the main panel respectively. $I_{\mathrm{pump}}$ is plotted in each case. The line scans exhibit plateaus at values corresponding to an integer number of electrons being transported per cycle of the RF signal. This is the signature of quantized charge transport.

The value of the current on the plateaus is proportional to the number of electrons $n_{e}$ ejected into the drain per cycle of the RF signal. Aspects of a model for the mechanism of operation of DyQD pumps have been discussed in previous works~\cite{blumenthal07,kaestner:153301,kaestner08}. In order to correctly interpret the results presented in this paper we present a detailed description of a model to explain the features seen in the pump map of Fig.~\ref{fig:pump_data}.

We draw the reader's attention to the four areas of the pump map along the entrance gate line scan (direction of constant $V_{\mathrm{exit}}$). These areas are labeled by the number of electrons captured and ejected in each case, $\left( n_{c},n_{e}\right)$. Schematic diagrams of the barriers defined by the entrance and exit gates in each area are presented in the right panel of Fig.~\ref{fig:pump_data}. Here, $E_{F}$ is the Fermi energy in the source (S) and drain (D) of the channel. In order to generate pumped current, it is necessary for $V_{\mathrm{exit}}$ to be negative enough at all points in the pump map for the barrier defined by the exit gate to always be opaque. The left and right schematics for each area represent the minimum and maximum barrier heights defined by the entrance gate during the pump cycle respectively. 

In area (0,0), the barrier defined by $V_{\mathrm{ent}}$ is too large over the whole pump cycle to allow electrons to enter the DyQD from the source. As the DyQD is never populated, we measure $I_{\mathrm{pump}}=0$ in this region.

As $V_{\mathrm{ent}}$ is made less negative the pump transitions into the (2,2) area where the entrance barrier drops enough to allow electrons to enter the DyQD from the source. When the entrance barrier subsequently rises as the pump cycle progresses we reach a point where the DyQD is isolated from the source. We refer to this point in the pump cycle as the \emph{capture point}, shown by the middle schematic. In this case the DyQD captures two electrons. By changing $V_{\mathrm{exit}}$, the size of the DyQD at the capture point can be altered and hence more or fewer electrons are captured. The captured electrons are subsequently ejected into the drain as the entrance barrier rises to its highest point. This results in a measured current in the (2,2) area of $I_{\mathrm{pump}}=2ef$. 

As $V_{\mathrm{ent}}$ becomes even less negative the pump switches to the (2,1) area where $I_{\mathrm{pump}}=ef$ is measured for the same exit gate voltage. The size of the DyQD at the capture point is expected to be the same as the previous case, so we assume that two electrons are again captured here but only one is ejected, with the other remaining confined within the DyQD. We therefore measure $I_{\mathrm{pump}}=ef$ in this area.

Finally, in area (2,0), the entrance barrier never rises high enough to push any of the captured electrons over the exit barrier and into the drain. The current measured in this region is therefore $I_{\mathrm{pump}}=0$.

The applied $V_{\mathrm{ent}}$ necessary for the entrance barrier to drop low enough to allow population of the dot (going from (0,0) to (2,2) in Fig.~\ref{fig:pump_data}) should be independent of $V_{\mathrm{exit}}$. The measured slope of the transition highlighted by the green dashed line in Fig.~\ref{fig:pump_data} arises from capacitive coupling between the gates.

We measure a different slope in the transitions corresponding to when $n_{e}$ differs from $n_{c}$. The transition from $n_{e}=n_{c}$ to $n_{e}=n_{c}-1$ (going from pumping all to pumping all but one electrons) is highlighted by the red dashed line in Fig.~\ref{fig:pump_data}. We believe this slope arises as a result of the shape of the potential at the stage in the pump cycle where the electrons are ejected into the drain being controlled by both $V_{\mathrm{ent}}$ and $V_{\mathrm{exit}}$.

\section{pumping in b$_{\perp}$}

\begin{figure}
\includegraphics[width=0.45\textwidth]{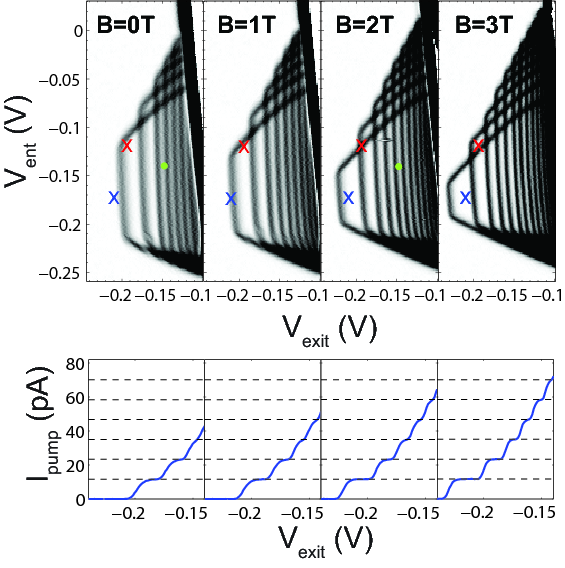}
\caption{\label{fig:B_dep} Upper panel: evolution of the pump map in the main left panel of Fig.~\ref{fig:pump_data} under the application of a perpendicular magnetic field $B_{\perp}$. Lower panel: line scans for $V_{\mathrm{ent}}=-0.17\,$V at each field increment. The dashed black lines mark the expected values for each plateau.}
\end{figure}

We next present data from measurements of the pumped current under the application of a perpendicular magnetic field to the device. We propose that information about the dynamics of the system may be extracted by monitoring changes in the pump map. Figure~\ref{fig:B_dep} shows the evolution of the pump map upon increasing $B_{\perp}$. The pumping frequency was set to $73\,$MHz and the amplitude of $V^{\mathrm{RF}}$ at the source was $-9\,$dBm, as before.

The upper panel of Fig.~\ref{fig:B_dep} shows that the transitions between plateaus become sharper (i.e. darker) as $B_{\perp}$ is increased. This suggests an enhancement of the quantization~\cite{wright08,kaestnerb}. The lower panel of Fig.~\ref{fig:B_dep} supplements this observation, where linescans in $V_{\mathrm{exit}}$ for $V_{\mathrm{ent}}=-0.17\,$V at each field increment are shown. It follows that the error mechanisms that give rise to deviations from perfectly quantized current at zero field must be suppressed at higher fields. A recent theoretical framework predicts the contribution of back-tunneling errors arising during the capture process~\cite{kaestner10}. In a perpendicular magnetic field we expect that the increased confinement of the captured electrons would lead to a reduction in the radial extent of the wave function~\cite{wright08}. We therefore expect a smaller overlap of the wave function with the leads and thus a lower probability of back-tunneling, resulting in an enhanced pumping accuracy.

The lower panel of Fig.~\ref{fig:B_dep} shows line scans in $V_{\mathrm{exit}}$ at each magnetic field for $V_{\mathrm{ent}}=-0.17\,$V. The plateaus are flatter in higher fields, as discussed. They are also longer, indicating enhanced robustness of the pumping mechanism~\cite{wright08}. In a field of $B_{\perp}=5\,$T these DyQD pumps, operating at $f=340\,$MHz, were shown to be accurate to better than 1.5 parts in 10$^{5}$~\cite{giblin10}. This result is important for quantum metrology and the development of a quantum standard for current. 

At higher fields we observe quantized current plateaus corresponding to a larger number of electrons robustly transported per cycle. The blue and red crosses in Fig.~\ref{fig:B_dep} are placed at the same coordinates in each plot, and they highlight the use of the field as an effective tuning parameter. In the case of the blue crosses the field is able to turn the pumping on in an area of the pump map where our model suggests that the dot is too small to capture electrons at 0~T. As illustrated in Fig.~\ref{fig:pump_data}, during the first part of the pump cycle the DyQD is coupled to the source, and so electrons are able to easily leave the DyQD and return to the source as the RF cycle progresses and the DyQD becomes smaller. The perpendicular magnetic field has the effect of increasing the effective confinement potential experienced by electrons in the DyQD, and so there is an enhanced probability of an electron remaining in the DyQD at the capture point. This explains the gradual increase in the pumped current from zero to $ef$ at the point indicated by the blue cross in Fig.~\ref{fig:B_dep} as $B_{\perp}$ is increased from 0~T to 3~T.

A similar explanation can be applied to the pumped current at the point indicated by the red cross in Fig.~\ref{fig:B_dep}. At 0~T, the red cross resides in the (1,1) region of the map, indicating that no electrons remain in the DyQD at the end of the pump cycle. Conversely, at a field of 3~T the red cross is in the (2,1) region. Here, we see that the increased confinement has led to a transition from capturing $n_c$ electrons to capturing $n_c+1$ electrons, as above, whilst also enabling the DyQD to confine a single electron at the end of the pump cycle ($n_c-n_e=0$ at 0~T, but $n_c-n_e=1$ at 3~T).

The green dots in the $B=0\,$T and 2~T pump maps of Fig.~\ref{fig:B_dep} serve to further illustrate this behaviour. In zero field the DyQD captures and ejects three electrons per cycle. Upon increasing the field to $B=2\,$T the DyQD was able to capture and eject five electrons for the same electrically defined DyQD. A full explanation of the evolution of the pump map in a magnetic field will require a more detailed computational study of electron dynamics in this device~\cite{adam}.




\section{pumping in high $B_{\perp}$}

\begin{figure}
\includegraphics[width=0.45\textwidth]{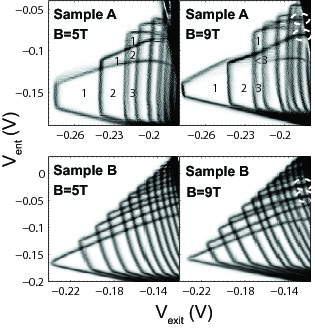}
\caption{\label{fig:high_b} Pump maps for large $B_{\perp}$. Lower panel: continuation of the data presented in Fig.~\ref{fig:B_dep}. The upper panel shows data collected with a different device, processed using a different HEMT wafer. $V^{\mathrm{RF}}$ for sample A was set to $f=306.7\,$MHz at an amplitude of $-9.6\,$dBm. Numbers in the plateaus correspond to the number of electrons transported per cycle in those regions.}
\end{figure}

At fields of $B>3\,$T the pump maps begin to exhibit phenomena that are no longer reproducible from device to device. Figure~\ref{fig:high_b} shows pumping maps at fields of $B=5\,$T and $9\,$T for two different samples. The data presented earlier in this work was collected with sample B. Sample A was fabricated using a different HEMT wafer and had a slightly different etched channel geometry. Sample A's pumping frequency was $f=306.7\,$MHz.

In each sample we observe an anomalous structure in the pumped current at high fields. For sample A the plateau corresponding to capturing two electrons and ejecting one electron (the last electron remaining confined within the DyQD at the end of the pump cycle) is no longer present at $9\,$T, as can be seen in the upper-right pump map of Fig.~\ref{fig:high_b}. The last two electrons appear to exit into the drain for the same entrance barrier height. Similar findings have been reported in SECS measurements where electrons were seen to tunnel into and out of static QDs in pairs and bunches over a range of $B_{\perp}$~\cite{ashoori1,zhitenev97}. Several theories which rely on disorder have been developed to explain this behavior~\cite{Wan95,Raikh96,canali99} but the origin remains unclear.

We did not observe identical behavior in sample B, but we did see other plateaus disappear at similar fields as transitions in the number of ejected electrons begin to merge. The white dashed ellipses in Fig.~\ref{fig:B_dep} highlight regions in the pump map where this merging can be observed. Different lines are seen to merge in each device. This behavior is also reminiscent of earlier SECS measurements where the addition spectra of different QDs displayed pairing and bunching of certain energy levels. In one experiment, artificial disorder was created by tuning the coupling of two nearby QDs. The pairing/bunching behavior was shown to be strongly dependent on the inter-dot coupling, and hence upon disorder~\cite{brodsky00}. Bunching behavior in our devices generally occurs for magnetic fields of at least $\sim5\,$T. In disordered systems it is expected that the field enhances disorder: the wave function shrinks, leading to an enhancement of the effects of a localization potential (for a review, see~\cite{pepper78}).


A full plateau structure persisted up to the maximum readily achievable fields in our measurement system of $15\,$T. Our results are very different from those published by Kaestner \emph{et al}.~\cite{kaestnerb}, where at 10.2$\,$T only one $n_{e}=1$ plateau was observed with all $n_{e}>1$ plateaus being completely suppressed. For certain frequencies, RF signal amplitudes and field strengths we did see similar patterns to those of Kaestner \emph{et al}. which we attribute to anomalous rectified biases that appear to be not only frequency dependent but also magnetic field dependent. The origin of rectification in our devices is not fully understood at present but is likely to be due to a complicated response of the sample holder, bond wires and ohmic contacts to the applied RF signal.

\section{conclusions}

In summary, we have presented experimental observations of the effect of a perpendicular magnetic field on the quantized current produced by DyQD electron pumps. The pumping accuracy was shown to be enhanced by the field, suggesting a suppression of the error mechanisms associated with a loss of quantization. The field was shown to be an effective extra control parameter in the tuning of the pump. As we increased the field to $B=3\,$T the pump could be turned on in a region of the pump map where no pumped current was generated at zero field. Our observations suggest the magnetic field is adding an extra confinement potential to the gate-defined DyQD. For $B>5\,$T we detected anomalous structure in the quantized current. We observed the onset of a pairing behavior in the ejected electrons reminiscent of SECS measurements, where several theories predict pair tunneling in QDs can arise from disorder unique to each individual QD. Our data suggests that local disorder, unique to each DyQD, affects the pumping more strongly for higher magnetic fields. We hope our findings will promote DyQDs as useful tools for probing few-electron dynamics in many fundamental investigations.

We gratefully acknowledge Bernd Kaestner, Christoph Leicht and Philipp Mirovsky for useful discussions. SJW acknowledges support from the EPSRC and Toshiba Research Europe Ltd. The work of MDB was supported by the UK National Measurement System’s Quantum Metrology Programme. The work of GG was supported by contract FA9453-07-C-0207 of AFRL. CAN acknowledges support from the EPSRC QIP IRC (GR/S82176/01).

\end{document}